\documentclass[aps,prd,preprint,floatfix]{revtex4} % preprint, % ,nofootinbib,twocolumn,showpacs,groupedaddress,

\usepackage{graphics}
\usepackage{graphicx}

\usepackage{latexsym}
\usepackage[cp866]{inputenc}
\usepackage[english]{babel} % ,russian

\input epsf

\begin{document}

\title{\boldmath
Observation of the isospin breaking decay
$\Upsilon(10860)\to\Upsilon(1S) f_0(980)\to\Upsilon(1S)\eta\pi^0$
with the Belle II detector}
\author{N.~N.~Achasov and G.~N.~Shestakov }
\affiliation{Laboratory of Theoretical Physics, S.~L.~Sobolev
Institute for Mathematics, 630090 Novosibirsk, Russia}

\begin{abstract} Using the Belle data on the decay $\Upsilon(10860)\to
\Upsilon(1S)f_0(980)\to\Upsilon(1S)\pi^+\pi^-$, we estimate the
fraction of the process $\Upsilon(10860)\to\Upsilon(1S)f_0(980)\to
\Upsilon(1S)\eta\pi^0$, caused by the mixing of $a^0_0(980)$ and
$f_0(980)$ resonances that breaks the isotopic invariance due to the
$K^+$ and $K^0$ meson mass difference. With an instantaneous
luminosity of $8\times10^{35}$ cm$^{-2}\cdot$\,s$^{-1}$ at the
SuperKEKB collider, one can collect with the Belle II detector about
a hundred $\Upsilon(10860)\to\Upsilon(1S)\eta\pi^0$ events in the
narrow region of the $\eta\pi^0$ invariant mass near the $K\bar K$
thresholds in a time roughly spent on the present Belle experiment.
\end{abstract}

\maketitle

Recently, the Belle Collaboration performed a full amplitude
analysis of three-body $e^+e^-\to\Upsilon(nS)\pi^+\pi^-$ ($n = 1, 2,
3$) transitions at $\sqrt{s}=10.865$ GeV and determined the relative
fractions of various quasitwo-body components of the three-body
amplitudes as well as the spin and parity of the two observed $Z_b$
states \cite{Ga15}. They also reported the first observation of the
$e^+e^-\to\Upsilon(1S)f_0(980)$ transition. The fraction of the
decay $\Upsilon(10860)\to\Upsilon(1S)f_0(980)$ is (see Table VI in
Ref. \cite{Ga15})
\begin{equation}\label{Eq1}
f_{\Upsilon(1S)f_0(980)}=\frac{BR(\Upsilon(10860)\to\Upsilon(1S)f_0
(980)\to\Upsilon(1S)\pi^+\pi^-)}{BR(\Upsilon(10860)\to\Upsilon(1S)
\pi^+\pi^-)}=\left(6.9\pm1.6^{+0.8}_{-2.8} \right)\%
\,.\end{equation}

\begin{figure}%[!ht]
\centerline{\epsfysize=7.cm    %%% 3.25cm
\epsfbox{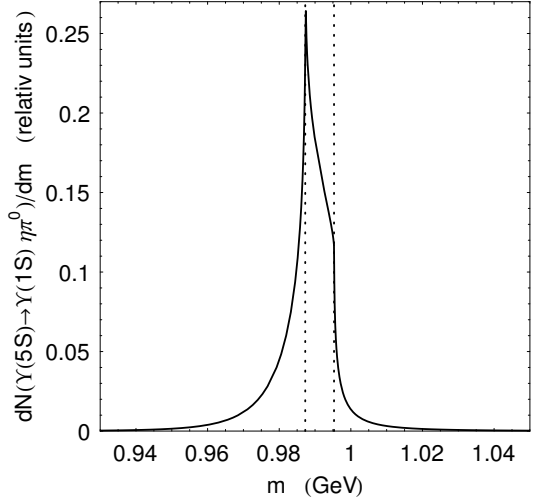}} %\vspace{-2mm} \hspace{-7mm}
\caption{\label{Figure1}The solid curve shows the $\eta\pi^0$ mass
spectrum in the decay $\Upsilon(10860)\to\Upsilon(1S)f_0(980)
\to\Upsilon (1S)\eta\pi^0$ calculated with the use of Eq.
(\ref{Eq4}). The dotted vertical lines show the locations of the
$K^+K^-$ and $K^0\bar K^0$ thresholds.} \label{Fig1}\end{figure}

\begin{figure}%[!ht]
\hspace*{0.3cm}\includegraphics[width=20pc]{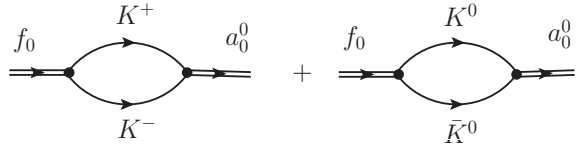}
\caption{\label{Figure2} The $K\bar K$ loop mechanism of the
$a^0_0(980)-f_0(980)$ mixing.}\end{figure}

The resonance $f_0(980)$ can decay into $\eta\pi^0$  via the
transition $f_0(980)\to(K^+K^-+K^0\bar K^0)\to a^0_0(980)\to\eta
\pi^0$, i.e., owing to the $a^0_0(980)-f_0(980)$ mixing
\cite{ADS79,AS17}. Using, as a guide, the central value for the
$a^0_0(980)-f_0(980)$ mixing intensity measured by the BESIII
Collaboration in the reaction $J/\psi\to\phi f_0(980) \to\phi
a^0_0(980)\to\phi\eta\pi^0$ \cite{Ab11},
\begin{equation}\label{Eq2}
\xi_{fa}=\frac{BR(f_0(980)\to K\bar K\to a^0_0(980)\to\eta\pi^0)}
{BR(f_0(980)\to\pi^+\pi^-)}\approx0.009,
\end{equation} we obtain the following estimate for the isospin breaking
decay fraction of the $\Upsilon(10860)$
\begin{equation}\label{Eq3} f_{\Upsilon(1S)a^0_0(980)}=
\frac{BR(\Upsilon(10860)\to\Upsilon(1S)f_0(980)\to\Upsilon(1S)\eta\pi^0)}
{BR(\Upsilon(10860)\to\Upsilon(1S)\pi^+\pi^-)}\approx6.2\cdot10^{-4}\,.
\end{equation}
The most characteristic feature of the decay $\Upsilon(10860)\to
\Upsilon(1S)f_0(980)\to\Upsilon(1S)\eta\pi^0$ is the dominance of
the narrow resonance structure in the $\eta\pi^0$ mass spectrum in
the vicinity of the $K\bar K$ thresholds \cite{AS17,AKS16}. The
corresponding $\eta\pi^0$ mass spectrum (see Fig. \ref{Figure1}) is
given by
\begin{eqnarray}\label{Eq4}
\frac{dN(\Upsilon(5S)\to\Upsilon(1S)\eta\pi^0)}{dm}= Cp(m)\left|
\frac{\Pi_{a^0_0f_0}(m)}{D_{a^0_0}(m)D_{f_0}(m)-\Pi^2_{a^0_0f_0}
(m)}\right|^2\frac{2m^2\Gamma_{a^0_0\to\eta\pi^0}(m)}{\pi},
\end{eqnarray}
where $\Upsilon(5S)$ is a short notation for $\Upsilon(10860)$, $m$
is the invariant mass of the $\eta\pi^0$ system,
$p(m)=\sqrt{m^4_{\Upsilon(5S)}-2m^2_{\Upsilon(5S)}(m^2+m^2_{\Upsilon(1S)
})+(m^2-m^2_{\Upsilon(1S)})^2}\,/(2m_{\Upsilon(5S)})$, and $C$ is
the normalization constant.
%----------------------------------------------------------------------------------
%----------------------------------------------------------------------------------
The $a^0_0(980)-f_0(980)$ mixing amplitude $\Pi_{a^0_0f_0}(m)$ in
Eq. (\ref{Eq4}), caused by the diagrams shown in Fig. \ref{Figure2},
has the form
%-------------------------------------------------------------------------------
%-------------------------------------------------------------------------------
\begin{eqnarray}\label{Eq5} %\hspace*{-35pt} &&
&& \Pi_{a^0_0f_0}(m)=\frac{g_{a^0_0K^+K^-}g_{f_0K^+K^-
}}{16\pi}\Biggl[\,i\,\Bigl(\rho_{K^+K^-}(m)%\nonumber\\ \hspace*{-35pt} &&
-\rho_{K^0\bar K^0}(m)\Bigr)\nonumber\\ && -
\frac{\rho_{K^+K^-}(m)}{\pi}\ln\frac{1+\rho_{K^+K^-}(m)}
{1-\rho_{K^+K^-}(m)}%\nonumber\\ \hspace*{-35pt} &&
+\frac{\rho_{K^0 \bar K^0}(m)}{\pi}\ln\frac{1+\rho_{K^0\bar
K^0}(m)}{1-\rho_{K^0\bar K^0}(m)}\,\,\Biggl]\,,\end{eqnarray} where
$m\geq2m_{K^0}$ and $\rho_{K\bar K}(m)=\sqrt{1-4m_K^2/m^2}$; in the
region $0\leq m\leq2m_{K^0}$, $\rho_{K^0\bar K^0}(m)$ should be
replaced by $i|\rho_{K^0\bar K^0}(m)|$, and in the region $0\leq
m\leq2m_{K^+}$, $\rho_{K^+K^-} (m)$ should be replaced by
$i|\rho_{K^+K^-}(m)|$. $D_r(m)$, in Eq. (\ref{Eq4}), is the inverse
propagator of the unmixed resonance $r$ $[r=a^0_0(980),f_0(980)]$
with the mass $m_r$\,,
\begin{eqnarray}\label{Eq6}
D_r(m)=m^2_r-m^2+\sum_{ab}[\mbox{Re}\Pi^{ab}_r(m_r)-\Pi^{ab}_r(m)],
\end{eqnarray} $ab=(\eta\pi^0,\,K^+K^-,\,K^0\bar K^0,\,\eta'\pi^0)$
for $r=a^0_0(980)$ and $ab=(\pi^+\pi^-,\,\pi^0\pi^0,\,K^+K^-,\,K^0
\bar K^0,\,\eta\eta)$ for $r=f_0(980)$; $\Pi^{ab}_r(m)$ stands for
the diagonal matrix element of the polarization operator of the
resonance $r$ corresponding to the contribution of the $ab$
intermediate state. At $m>m_a+m_b$,
\begin{eqnarray}\label{Eq7}\Pi^{ab}_{r}(m)=\frac{g^2_{r
ab}}{16\pi} \left[\frac{m_{ab}^{(+)}m_{ab}^{(-)}}{\pi
m^2}\ln\frac{m_b}{m_a}+\rho_{ab}(m)%\right.\ \nonumber\\ \left.\times
\left(i-\frac{1}{\pi}\,\ln\frac{\sqrt{m^2-m_{ab}^{(-)
\,2}}+\sqrt{m^2-m_{ab}^{(+)\,2}}}{\sqrt{m^2-m_{ab}^{(-)\,2}}-\sqrt{m^2
-m_{ab}^{(+)\,2}}}\right)\right],\end{eqnarray} where $g_{rab}$ is
the coupling constant of $r$ with $ab$, $\rho_{ab}(m)$\,=\,$
\sqrt{m^2-m_{ab}^{(+)\,2}} \,\sqrt{m^2-m_{ab}^{(-)\,2}}\,/m^2$,
$m_{ab}^{(\pm)}$\,=\,$m_a\pm m_b$, and $m_a\geq m_b$; $\mbox{Im}
\,\Pi^{ab}_r(m)=m\Gamma_{r\to ab}(m)=(g^2_{r
ab}/16\pi)\rho_{ab}(m)$. At $m_{ab}^{(-)}<m<m_{ab}^{(+)}$
\begin{eqnarray}\label{Eq8}\Pi^{ab}_{r}(m)=\frac{g^2_{r
ab}}{16\pi} \left[\frac{m_{ab}^{(+)}m_{ab}^{(-)}}{\pi
m^2}\ln\frac{m_b}{m_a}%\right.\nonumber\\ \left.
-\rho_{ab}(m)\left(1-\frac{2}{\pi}\arctan\frac{\sqrt{
m_{ab}^{(+)\,2}-m^2}}{\sqrt{m^2-m_{ab}^{(-)\,2}}}\right)\right],
\end{eqnarray}
where  $\rho_{ab}(m)$\,=\,$\sqrt{m_{ab}^{(+)\,2}-m^2}
\,\sqrt{m^2-m_{ab}^{(-)\,2}}\,/m^2$. At $m\leq m_{ab}^{(-)}$
\begin{eqnarray}\label{Eq9}\Pi^{ab}_{r}(m)=\frac{g^2_{r
ab}}{16\pi} \left[\frac{m_{ab}^{(+)}m_{ab}^{(-)}}{\pi
m^2}\ln\frac{m_b}{m_a}%\right.\ \ \nonumber\\ \left.
+\rho_{ab}(m)\frac{1}{\pi}\,\ln\frac{
\sqrt{m_{ab}^{(+)\,2}-m^2}+\sqrt{m_{ab}^{(-)\,2}-m^2}}
{\sqrt{m_{ab}^{(+)\,2}-m^2}-\sqrt{m_{ab}^{(-)\,2}-m^2}}\right],
\end{eqnarray}
where $\rho_{ab}(m)$\,=\,$\sqrt{m_{ab}^{(+)\,2}-m^2}\,\sqrt{
m_{ab}^{(-)\,2}-m^2}\,/m^2$. Here we use as a guide the numerical
estimates of the coupling constants $g^2_{f_0 ab}/(16\pi)$ and
$g^2_{a^0_0ab}/(16\pi)$ obtained in Ref. \cite{AKS16} by analyzing
the BESIII data on the $a^0_0(980)-f_0(980)$ mixing \cite{Ab11}
\begin{eqnarray}\label{Eq10}
\frac{g^2_{f_0\pi\pi}}{16\pi}\equiv\frac{3}{2}\frac{g^2_{f_0\pi^+\pi^-}}
{16\pi}=0.098\mbox{\ GeV}^2,\\ \label{Eq11} \frac{g^2_{f_0 K\bar
K}}{16\pi}\equiv2\frac{g^2_{f_0 K^+K^-}}{16\pi}=0.4\mbox{\ GeV}^2,
\\ \label{Eq12} \frac{g^2_{a^0_0\eta\pi^0}}{16 \pi}=0.2\mbox{\ GeV}^2,
\quad\quad\  \\ \label{Eq13} \frac{g^2_{a^0_0 K\bar K}}{16\pi}
\equiv2\frac{g^2_{a^0_0 K^+K^-}}{16\pi}=0.5\mbox{\ GeV}^2.
\end{eqnarray}
As in Ref. \cite{AKS16}, we fix $m_{a^0_0}=0.985$ GeV, $m_{f_0}=
0.985 $ GeV (see also Ref. \cite{Fn1}) and set
$g^2_{a^0_0\eta'\pi^0}=g^2_{a^0_0\eta\pi^0}$ and
$g^2_{f_0\eta\eta}=g^2_{f_0 K^+K^-}$ by the $q^2\bar q^2$ model.

%--------------------------------------------------------------------------------------

In the Belle experiment of 2015 \cite{Ga15}, $2090\pm115$
$\Upsilon(10860)\to\Upsilon(1S)\pi^+\pi^-$ events were collected.
Thus, one or two $\Upsilon (10860)\to\Upsilon(1S)\eta\pi^0$ events
could be produced in this experiment due to the $a_0(980)-f_0(980)$
mixing.

There is no visible $\eta\pi^0$ background and one can hope that
this rare decay will be measured with the use of the Belle II
detector at the SuperKEKB with at least 10\% statistical accuracy.
In the new experiment with a 40 times greater instantaneous
luminosity, one can collect about a hundred $\Upsilon (10860)\to
\Upsilon(1S)\eta\pi^0$ events (i.e., from 50 to 150 events) in the
narrow region of the $\eta\pi^0 $ invariant mass near the $K\bar K$
thresholds in the time comparable with that spent on the initial
Belle experiment \cite{Ga15}. According to our estimates, the
statistics in the experiment \cite{Ga15} were collected in about
three months.

The observation of the decay $\Upsilon (10860)\to\Upsilon(1S)f_0
(980)\to\Upsilon(1S)\eta\pi^0$, together with investigations of the
weak hadronic decays $D_s^+\to\eta\pi^0\pi^+$ \cite{AS17a}, $D^0\to
K_S\pi^+\pi^-$ and $D^0\to K_S\eta\pi^0$ \cite{AS17b}, can open a
new stage in a thorough study of the $a^0_0(980)-f_0(980)$ mixing
phenomenon and the nature of the light scalar mesons.

We are grateful to A.E. Bondar, who drew our attention to the decay
under consideration. The present work is partially supported by the
Russian Foundation for Basic Research Grant No. 16-02-00065 and the
Presidium of the Russian Academy of Sciences Project Grant No.
0314-2015-0011.

%------------------------------------------------------------------------------------------------------------

\end{document}